# Direct Power Flow Controller with Continuous Full Regulation Range

Yao Chong, Youjun Zhang

***Abstract*—For enhancing power flow control in power transmission, a simplified new structure of direct power flow controller with continuous full regulation range (F-DPFC) was proposed. It has only one-stage power conversion and comprises of a three-phase transformer in parallel and a three-phase transformer in series with grid, three single-phase full-bridge ac units, and a three-phase filter. Compared with previous DPFC, the proposed one dispenses with two complex three-phase selection switches which connect with high-voltage grid directly, and has a continuous 360° adjustment range of compensation voltage by taking place of buck-type ac unit with full-bridge type ac unit, and then expanding the limit of its duty cycle from [0,1] to [-1,1]. Within a large smooth zone replacing six separate zones, the proposed F-DPFC can regulate the amplitude and phase angle of grid node voltage respectively and simultaneously, and then the active and reactive power flow in grid can be controlled smoothly and effectively. The new structure is easy to achieve modular expansion and enables it to operate under high voltage and power conditions. Its structure and operational principle were analyzed in detail, and a prototype was developed. The experimental results verified the feasibility and the correctness of the theoretical analysis.**

## I. INTRODUCTION

How to control the power flow fast and accurately in power transmission systems is always the key to improve power grid quality and energy transfer efficiency. Power flow in power system includes active power flow and reactive power flow which are determined by the line impedance, transmission angle, and bus voltage [1]. Therefore, the flexible ac-transmission system (FACTS) which can adjust one or more ac-transmission system parameters to increase the stability of power system is widely used [2-4], among which the unified power flow controller (UPFC) [5-6] is the most common.

The UPFC consists of a static synchronous compensator (STATCOM) [7,8] and a static synchronous series compensator (SSSC) [7,9], which are connected through a large dc energy storage capacitor. STATCOM which is one of shunt FACTS devices [10] can regulate reactive power by being connected in parallel to the power transmission, and SSSC is able to control the line current and active power flow as series FACTS devices [11]. By combining STATCOM and SSSC, UPFC can realize the function of adjusting active and reactive respectively and simultaneously. However, due to the large dc energy storage element between STATCOM and SSSC, which has short equipment life cycle or a high number of failures and cause high maintenance cost, the further promotion of UPFC is restricted, even though it has superior control ability.

Accordingly, a three-phase power flower controller with direct PWM AC/AC converters [12] was presented, which is able to regulate the amplitude and phase of voltage without a dc energy storage element. It uses the vector synthesis of two-phase voltage to achieve the function of regulating voltage and adopts the circuit of bipolar matrix chopper which is more functional than quadrature shifter [13] to improve the ability of regulation. But this also leads to their need for complex structure and control system, and the circuit has low energy transfer efficiency. Moreover, its circuit structure is not easy to achieve modular expansion [14,15] and difficult to apply in high-voltage and high-power applications.

A new concept called direct power flow controller (DPFC) which does not contain large dc energy storage elements either was described in [4]. As shown in Fig.1, DPFC is based on single-stage ACCPA [16] and it replaces the boost-type ac converter in two-stage ACCPA [17-19] with an output transformer to achieve the function of boosting voltage and has only one-stage conversion circuit. DPFC is able to regulate the amplitude and phase angle of output compensation voltage respectively and simultaneously within 360° range, and adjust active power and reactive power in power transmission system by connecting the out compensation voltage in series to the power grid.

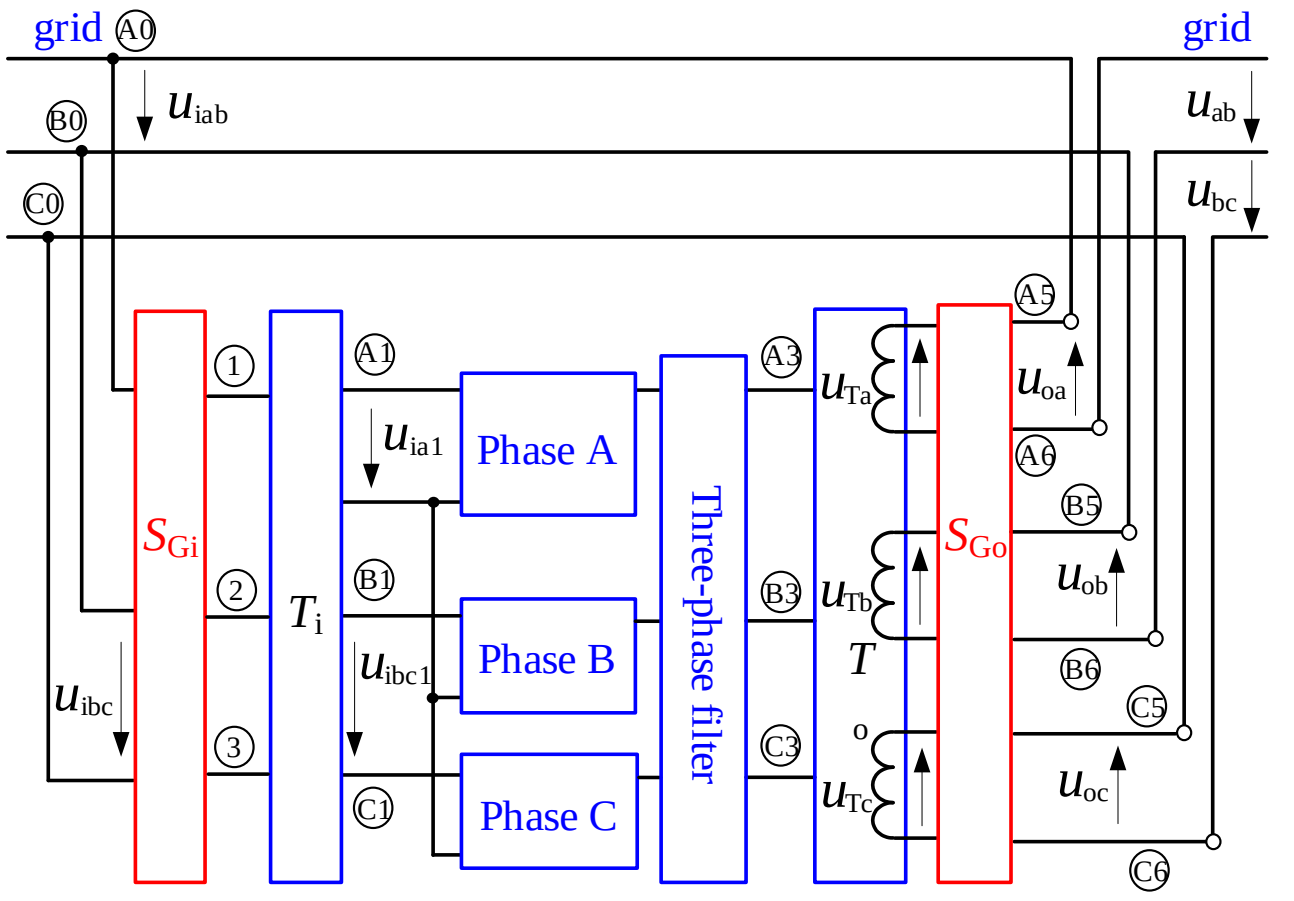

Fig. 1. DPFC with full 360° regulation zone.

The regulation range of output compensation voltage in DPFC is shown in Fig.2, and we can see it is divided into six separate zones. Under a combination of selector switches, the adjustment range of the output compensation voltage phase angle is 60°, and only with six combinations can the phase angle adjustment range be extended to 360°. The adjustment process is only continuous within a separate zone and only by changing the combination of the select switches can the output compensation voltage be adjusted from one zone to another which is apparently not conducive to the stability of power transmission.

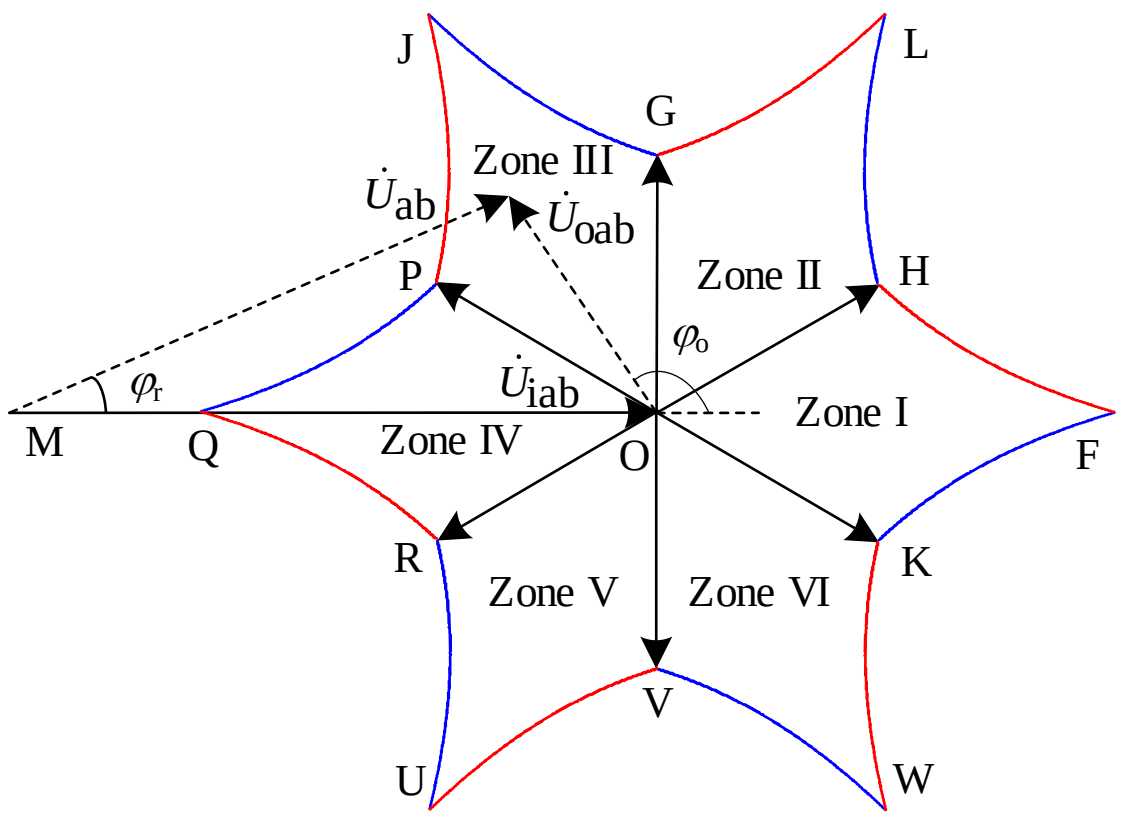

Fig. 2. Six basic regulation zones of DPFC.

In order to solve the problem of discontinuous control processes in DPFC and simplify structure for modular expansion, a new structure of direct power flow controller with continuous full regulation range (F-DPFC) was proposed in this paper. F-DPFC replaces buck-type ac units in DPFC with full-bridge type ac units and removes selector switches which have complex structure and high cost owing to being connected to the high voltage power grid directly. F-DPFC also does not contain dc energy storage and has one-stage conversion circuit. Without the help of selection switches, F-DPFC can output a compensation voltage whose phase angle can vary within 360°, and then adjust the amplitude and phase angle of grid node voltage individually or simultaneously to control active power flow and reactive power flow in power transmission system. The topology structure and operational principle of F-DPFC were described in detail. The regulation range of output compensation voltage and relationship between adjustment range and control parameters were analyzed, and then the selection of control parameters and closed-loop control strategy were given. Finally, a prototype of F-DPFC and experiment result were shown to verify the correctness of theory and feasibility of F-DPFC.

## II. TOPOLOGY SRUCTURE AND OPERATIONAL PRINCIPLE

### *A. Topology Structure*

The topology structure of F-DPFC is shown in Fig. 3. Similar to DPFC, the structure of F-DPFC also contains a three-phase transformer in parallel and a three-phase transformer in series with grid. It takes place of buck type ac units in DPFC with full-bridge type ac units and removes selection switches. The circuit structure is further simplified and the circuit cost is reduced.

Fig. 4 shows the connection mode of input transformer winding and output transformer winding. The input terminal of transformer $T_i$ is connected in parallel with the power grid and the output terminal of the transformer $T_o$ is connected in series with the power grid. In this paper, $T_i$ and $T_o$ are of Δ/Yn11 type connection group to offset third harmonic current and voltage. Each secondary winding of transformer $T_i$ is connected to the input terminal of full-bridge type ac units. The output terminals F (as shown in Fig. 3) of full-bridge type ac units are connected together and another output terminals are connected to the input terminal of transformer $T_o$ through three-phase output filter.

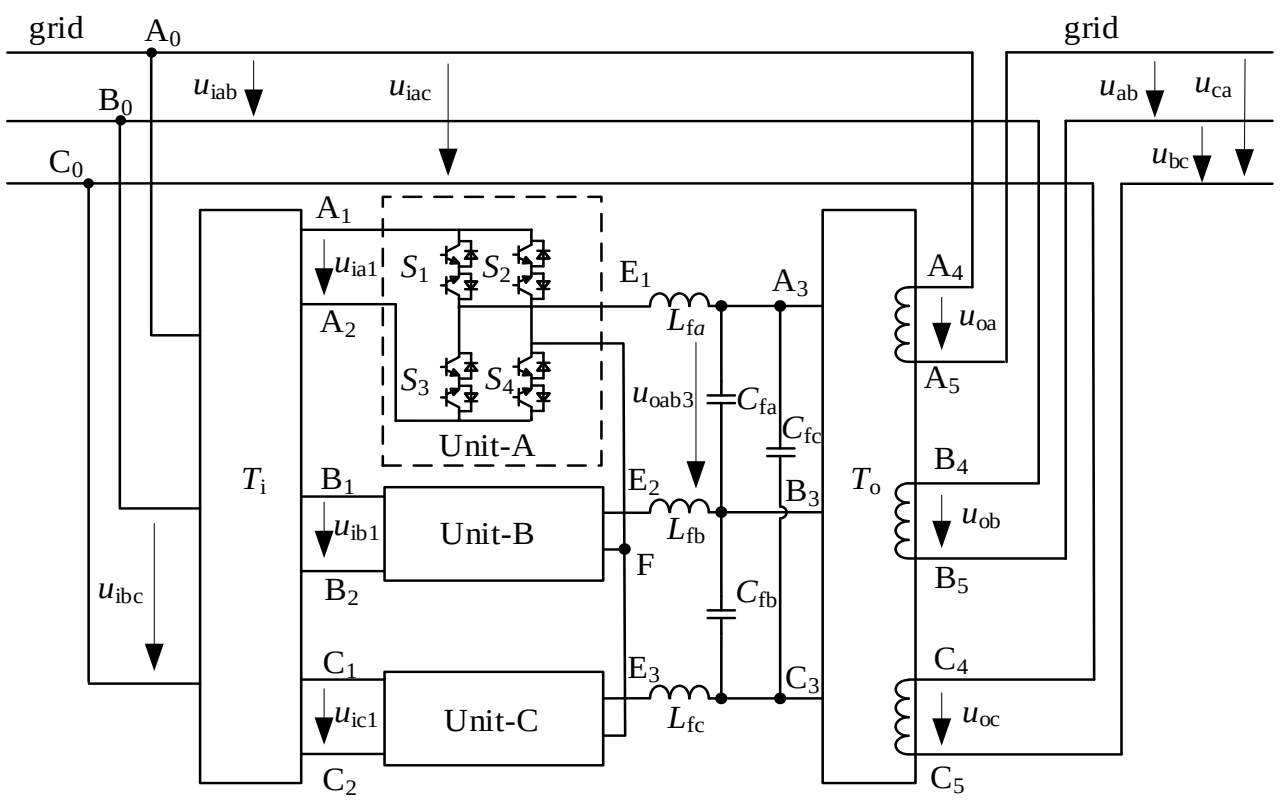

Fig. 3. Basic topology structure of the F-DPFC.

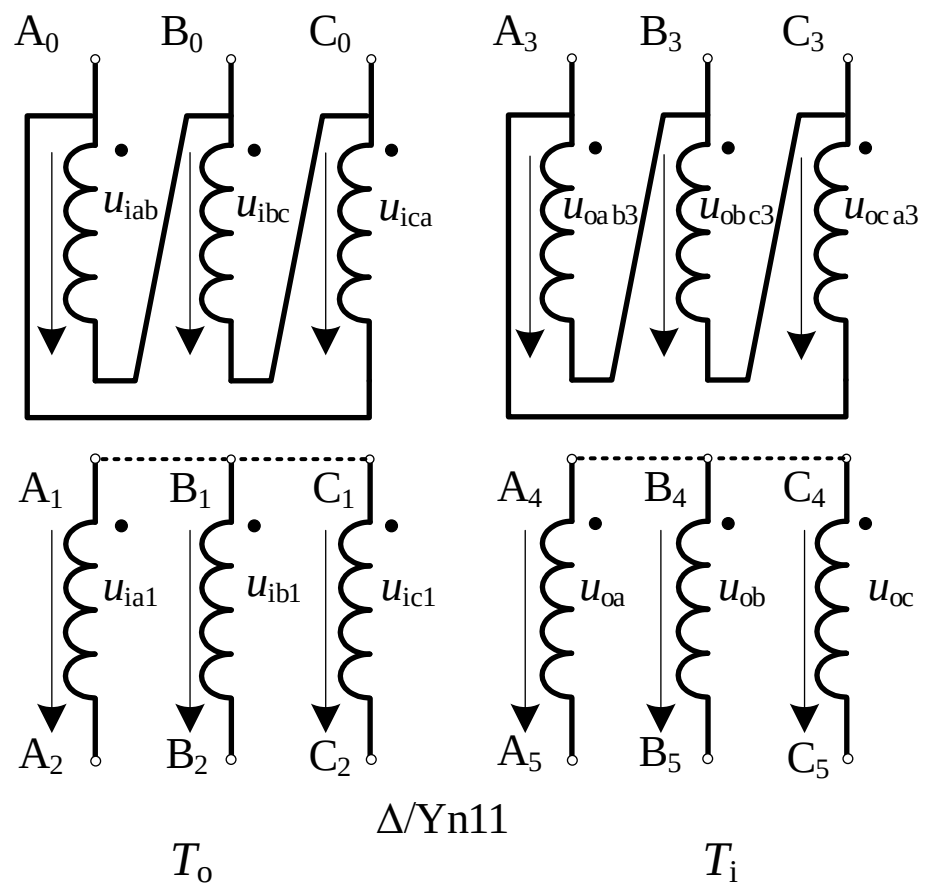

Fig. 4. The connection mode of input transformer and output transformer.

### *B. Operational Principle*

To facilitate analysis, we assume that:

1) original grid voltage is sinusoidal with angular frequency $\omega$ (=$2\pi f$, where $f$ is its frequency) and the original line voltage is $u_{iab}$, $u_{ibc}$ and $u_{ica}$.

2) circuit components are ideal and low-frequency voltage drop across the inductor $L_{fx}$ (x=a, b, c, where x is the name of phase in lowercase letter) is not taken into account.

3) $T_i$ and $T_o$ are of Δ/Yn11 type connection group with turn ratio $N_i$ and $N_o$ (note that points $A_1$, $B_1$ and $C_1$ or $A_4$, $B_4$ and $C_4$ are not connected together, and the secondary windings of $T_o$ are separately connected with power transmission line in series).

The input voltages $u_{ia1}$, $u_{ib1}$, $u_{ic1}$ and the duty ratio $d_a$, $d_b$, $d_c$ of power flow control units-A, B, C are as below:

$$\begin{cases} u_{ia1} = \dfrac{u_{iab}}{N_i} = \dfrac{U_{imL}}{N_i}\sin\omega t = U_{im}\sin\omega t \\ u_{ib1} = \dfrac{u_{ibc}}{N_i} = \dfrac{U_{imL}}{N_i}\sin(\omega t - 120°) = U_{im}\sin(\omega t - 120°) \\ u_{ic1} = \dfrac{u_{ica}}{N_i} = \dfrac{U_{imL}}{N_i}\sin(\omega t + 120°) = U_{im}\sin(\omega t + 120°) \end{cases} \quad (1)$$

$$\begin{cases} d_a = k_0 + k_2 \sin(2\omega t + \beta_2) \\ d_b = k_0 + k_2 \sin[2(\omega t - 120°) + \beta_2] \\ d_c = k_0 + k_2 \sin[2(\omega t + 120°) + \beta_2] \end{cases} \quad (2)$$

where $U_{imL}$ and $U_{im}$ are the amplitude of $u_{iab}$ and $u_{ia}$ respectively, $\beta_2$ is the initial phase angle of the ac component of $d_a$, and parameter $k_2$ is nonnegative. In unit-A in Fig. 3, when switch unit $S_1$ is always on and switch unit $S_3$ is always off and switch units $S_2$ and $S_4$ are alternately on, $0<d_a<1$; when switch unit $S_3$ is always on and switch unit $S_1$ is always off and switch units $S_2$ and $S_4$ are alternately on, $-1<d_a<0$. Therefore, we can get $-1 \le d_a \le 1$ and further know that $k_0+k_2 \le 1$ at $0 \le k_0 \le 1$, or $k_0-k_2 \ge -1$ at $-1 \le k_0 \le 0$. The value range of $k_0$, $k_2$ is shown in Fig. 5.

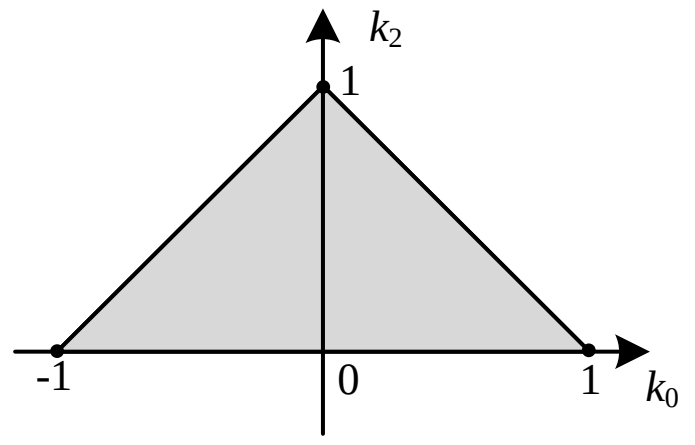

Fig. 5. Value range of $k_0$ and $k_2$.

After output filter filters out the high-frequency component, $u_{ox2}$ (between point $X_3$ and F, not shown in Fig. 3, X=A, B, C, X is the name of phase in uppercase letter) which are the output voltages of full-bridge type ac units are obtained:

$$\begin{cases} u_{oa2} = d_a u_{ia} \\ \quad = U_{im} \sin\omega t[k_0 + k_2 \sin(2\omega t + \beta_2)] \\ u_{ob2} = d_b u_{ib} \\ \quad = U_{im} \sin(\omega t - 120°)[k_0 + k_2 \sin(2\omega t + 120° + \beta_2)] \\ u_{oc2} = d_c u_{ic} \\ \quad = U_{im} \sin(\omega t + 120°)[k_0 + k_2 \sin(2\omega t - 120° + \beta_2)] \end{cases} \quad (3)$$

Simplified:

$$\begin{cases} u_{oa2} = U_{im}[k_0 \sin\omega t + \frac{1}{2}k_2 \sin(\omega t + \beta_2 + 90°) \\ \quad - \frac{1}{2}k_2 \cos(3\omega t + \beta_2)] \\ u_{ob2} = U_{im}[k_0 \sin(\omega t - 120°) + \frac{1}{2}k_2 \sin(\omega t + \beta_2 - 30°) \\ \quad - \frac{1}{2}k_2 \cos(3\omega t + \beta_2)] \\ u_{oc2} = U_{im}[k_0 \sin(\omega t + 120°) + \frac{1}{2}k_2 \sin(\omega t + \beta_2 + 210°) \\ \quad - \frac{1}{2}k_2 \cos(3\omega t + \beta_2)] \end{cases} \quad (4)$$

If $T_o$ is of Δ/Y11 type connection group with turn ratio $N_o$, the third harmonic signals in $u_{oa2}$, $u_{ob2}$ and $u_{oc2}$ cancel each other out, and only the fundamental voltage components $u_{oa3}$, $u_{ob3}$, $u_{oc3}$ are remained. One can obtain the voltages $u_{oab3}$, $u_{obc3}$, $u_{oca3}$.

$$\begin{cases} u_{oa3} = U_{im}[k_0 \sin\omega t + \frac{1}{2}k_2 \sin(\omega t + \beta_2 + 90°)] \\ \quad = U_{om} \sin(\omega t + \varphi_1) \\ u_{ob3} = U_{im}[k_0 \sin(\omega t - 120°) + \frac{1}{2}k_2 \sin(\omega t + \beta_2 - 30°)] \\ \quad = U_{om} \sin(\omega t + \varphi_1 - 120°) \\ u_{oc3} = U_{im}[k_0 \sin(\omega t + 120°) + \frac{1}{2}k_2 \sin(\omega t + \beta_2 + 210°)] \\ \quad = U_{om} \sin(\omega t + \varphi_1 + 120°) \end{cases} \quad (5)$$

$$\begin{cases} u_{oab3} = u_{oa3} - u_{ob3} = \sqrt{3}U_{om} \sin(\omega t + \varphi_1 + 30°) \\ u_{obc3} = u_{ob3} - u_{oc3} = \sqrt{3}U_{om} \sin(\omega t + \varphi_1 - 90°) \\ u_{oca3} = u_{oc3} - u_{oa3} = \sqrt{3}U_{om} \sin(\omega t + \varphi_1 + 150°) \end{cases} \quad (6)$$

where $U_{om}$ is the amplitude of the fundamental voltage component $u_{oa3}$, $\varphi_1$ is the phase angle of the fundamental wave voltage component $u_{oa3}$.

Here $U_{om}$ and $\varphi_1$ would be given as:

$$U_{om} = U_{im}\sqrt{\frac{k_2^{\ 2}}{4} + k_0^{\ 2} - k_0 k_2 \sin\beta_2} \quad (7)$$

$$\varphi_1 = \arctan\frac{k_2 \cos\beta_2}{2k_0 - k_2 \sin\beta_2} \quad (8)$$

Then one can obtain the compensation phase voltages of grid $u_{oa}$, $u_{ob}$, $u_{oc}$ and the compensation line voltages of grid $u_{oab}$, $u_{obc}$, $u_{oca}$:

$$\begin{cases} u_{oa} = \frac{\sqrt{3}U_{om}}{N_o} \sin(\omega t + \varphi_1 + 30°) \\ u_{ob} = \frac{\sqrt{3}U_{om}}{N_o} \sin(\omega t + \varphi_1 - 90°) \\ u_{oc} = \frac{\sqrt{3}U_{om}}{N_o} \sin(\omega t + \varphi_1 + 150°) \end{cases} \quad (9)$$

$$\begin{cases} u_{oab} = u_{oa} - u_{ob} = \frac{3U_{om}}{N_o} \sin(\omega t + \varphi_1 + 60°) \\ u_{obc} = u_{ob} - u_{oc} = \frac{3U_{om}}{N_o} \sin(\omega t + \varphi_1 - 60°) \\ u_{oca} = u_{oc} - u_{oa} = \frac{3U_{om}}{N_o} \sin(\omega t + \varphi_1 + 180°) \end{cases} \quad (10)$$

The regulated grid line voltages ($u_{ab}$, $u_{bc}$ and $u_{ca}$) are as follows:

$$\begin{cases} u_{ab} = u_{iab} + (u_{oa} - u_{ob}) = U_{mL} \sin\left(\omega t + \varphi_r\right) \\ u_{bc} = u_{ibc} + (u_{ob} - u_{oc}) = U_{mL} \sin\left(\omega t + \varphi_r - 120°\right) \\ u_{ca} = u_{ica} + (u_{oc} - u_{oa}) = U_{mL} \sin\left(\omega t + \varphi_r + 120°\right) \end{cases} \quad (11)$$

where $U_{mL}$ and $\varphi_r$ are the amplitude and phase angle of $u_{ab}$,

and here $U_{mL}$ and $\varphi_r$ would be as:

$$\varphi_r = \arctan \frac{\frac{3U_{imL}}{N_i N_o} \sin(\varphi_1 + 60°)}{U_{imL} + \frac{3U_{imL}}{N_i N_o} \cos(\varphi_1 + 60°)} \tag{12}$$

$$U_{mL} = \sqrt{[U_{imL} + \frac{3U_{imL}}{N_i N_o} \cos(\varphi_1 + 60°)]^2 + [\frac{3U_{imL}}{N_i N_o} \sin(\varphi_1 + 60°)]^2} \tag{13}$$

From (7), (8) we know that $U_{om}$ and $\varphi_1$ are controlled by three parameters ($k_0$, $k_2$ and $\beta_2$).

*C. Expansion of Circuit Modular Structure*

Due to material constraints, the voltage stress of power electronic device cannot be significantly increased. Therefore, modular expansion is an effective way for circuit structures to be suitable for high-voltage and high-power application. The structure of modular expansion is briefly introduced here.

Increase the number of the secondary windings of $T_i$ and the full-bridge ac circuits of each phase. Output terminals of the full-bridge ac circuits connected in series and the input terminals connected to the secondary windings of the three-phase multi winding transformer respectively. A-phase conversion unit of modular structure is shown in Fig. 6.

By increasing the number of modules, the maximum voltage stress that the circuit can withstand can be multiplied. The modular control strategy is flexible, and due to limited space, detailed explanations will be provided in the future.

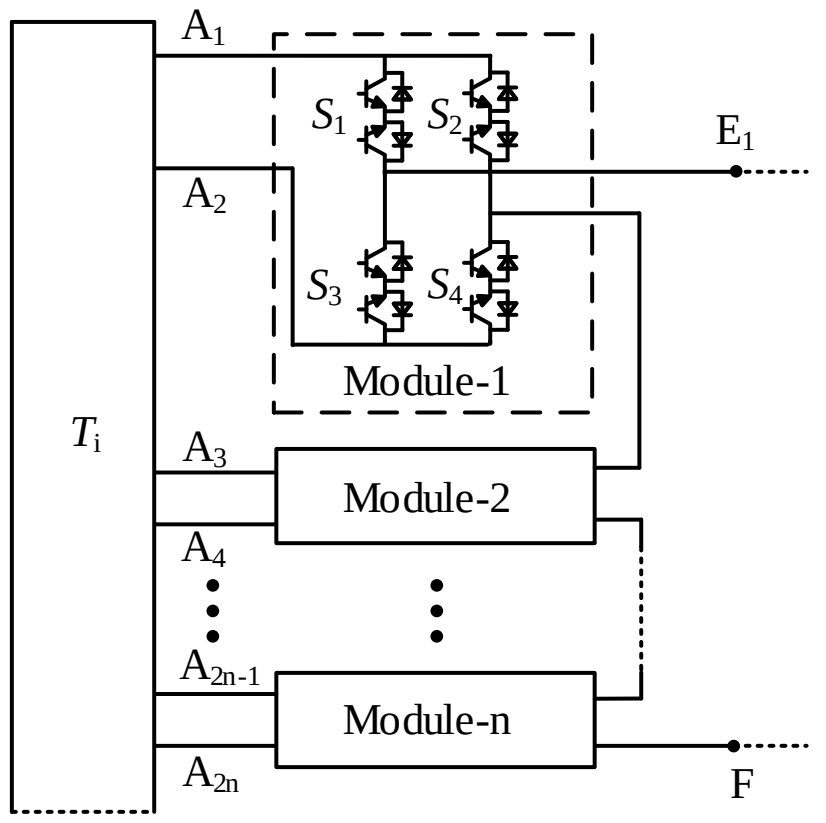

Fig. 6. Structure of phase-A modular expansion.

## III. ADJUSTMENT RANGE AND CONTROL STRATEGY

*A. Range of Grid Compensation Voltage*

For the convenience of analysis, take A-phase as an example. We can simplify the A-phase in formula (5) to the following formula (where $\beta=\beta_2+90°$):

$$k_d = \frac{u_{oa3}}{U_{im}} = k_0 \sin \omega t + \frac{1}{2} k_2 \sin(\omega t + \beta) \tag{14}$$

where $k_d$ is a vector with the same phase as $u_{oa3}$.

The compensation voltage synthesis diagram is shown in Fig. 7. In the diagram, the vector $OC$ represents $u_{oa3}/U_{im}$; the vector $OA$ represents the voltage component $k_0\sin\omega t$ and the vector $AC$ represents the voltage component $0.5k_2\sin(\omega t+\beta)$. The following can be obtained :

$$\begin{cases} |\overrightarrow{AB}| = \cos\beta \cdot |\overrightarrow{AC}| = \frac{k_2 \cos\beta}{2} \\ |\overrightarrow{BC}| = \sin\beta \cdot |\overrightarrow{AC}| = \frac{k_2 \sin\beta}{2} \\ |\overrightarrow{OB}| = |\overrightarrow{OA}| + |\overrightarrow{AB}| = k_0 + \frac{k_2 \cos\beta}{2} \end{cases} \tag{15}$$

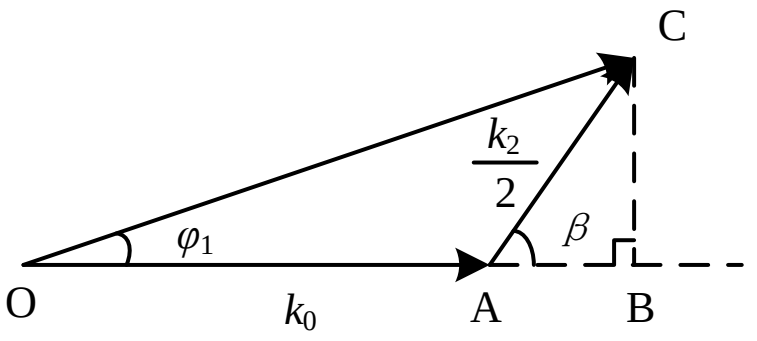

Fig. 7. Schematic diagram of compensation voltage synthesis.

Further obtain amplitude ratio $|k_d|$ and phase $\varphi_1$:

$$\begin{aligned} |k_d| &= |\overrightarrow{OC}| = \sqrt{|\overrightarrow{OB}|^2 + |\overrightarrow{BC}|^2} \\ &= \sqrt{(k_0 + \frac{k_2 \cos\beta}{2})^2 + (\frac{k_2 \sin\beta}{2})^2} \\ &= \sqrt{\frac{k_2^{\ 2}}{4} + k_0^{\ 2} + k_0 k_2 \cos\beta} \end{aligned} \tag{16}$$

$$\begin{aligned} \varphi_1 &= \arctan \frac{|\overrightarrow{BC}|}{|\overrightarrow{OB}|} = \arctan \frac{|\overrightarrow{BC}|}{|\overrightarrow{OA}| + |\overrightarrow{AB}|} \\ &= \arctan \frac{k_2 \sin\beta}{2k_0 + k_2 \cos\beta} \end{aligned} \tag{17}$$

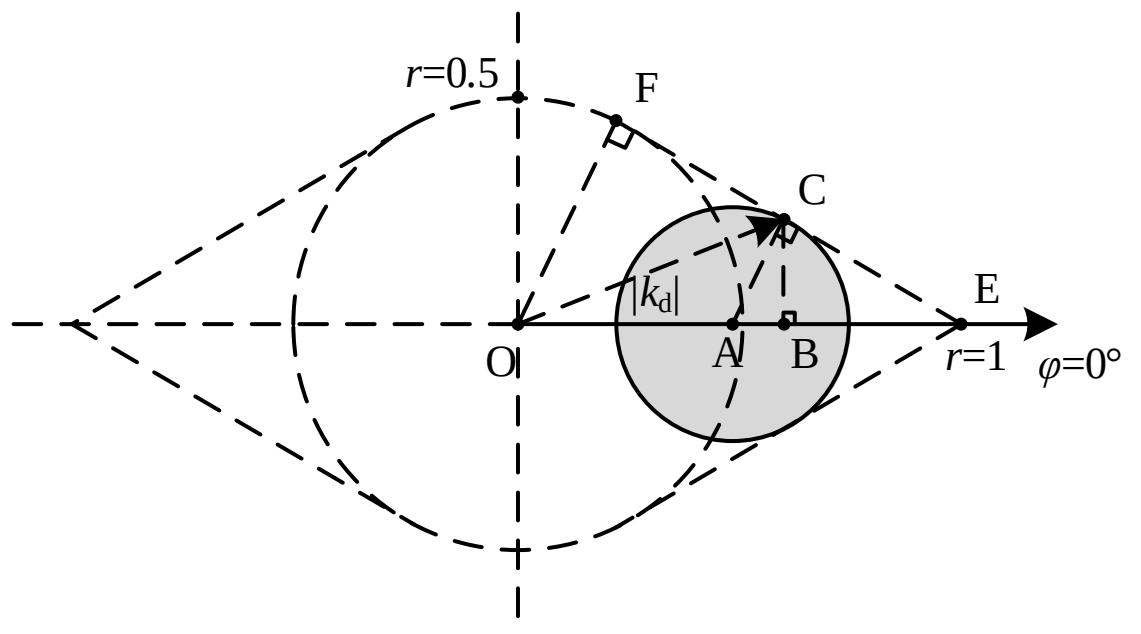

Fig. 8. When $k_0$ is a fixed value, the adjustment range of compensation voltage.

Combining the value range of $k_0$ and $k_2$, we can further obtain the total adjustment range of compensation voltage. Establish polar coordinates as shown in Fig. 8. Obviously, when $k_0=0$, the adjustment range is a circle with point O as the centre and 0.5 as the radius. When $k_0=1$, the adjustment range of compensation voltage is point E (1, 0°). Make the tangent line of circle O through point E and the tangent point is F, $\angle EOF=60°$. When $0<k_0<1$, assuming $OA=k_0$, as shown in Fig. 9, the adjustment range of the compensation voltage is a circle with point A as the centre, AC as the radius. One knows that $AC_{max}=0.5(1-k_0)$ and $AE=1-k_0$. In the right triangle ACE, $\angle ACE=90°$, $2AC=AE$. One can obtain $\angle CEA=30°$ and $\angle$FEO=30°, so points F, C and E are on the same straight line. The adjustment range of compensation voltage for other zones can be obtained in the same way. If $T_i$ and $T_o$ are of Δ/Yn11 type connection group, the total compensation voltage adjustment range as shown in Fig. 9 is obtained. In addition, under different connection group of $T_i$ and $T_o$, up to three different compensation adjustment ranges can be obtained, and phase difference among them is 60°.

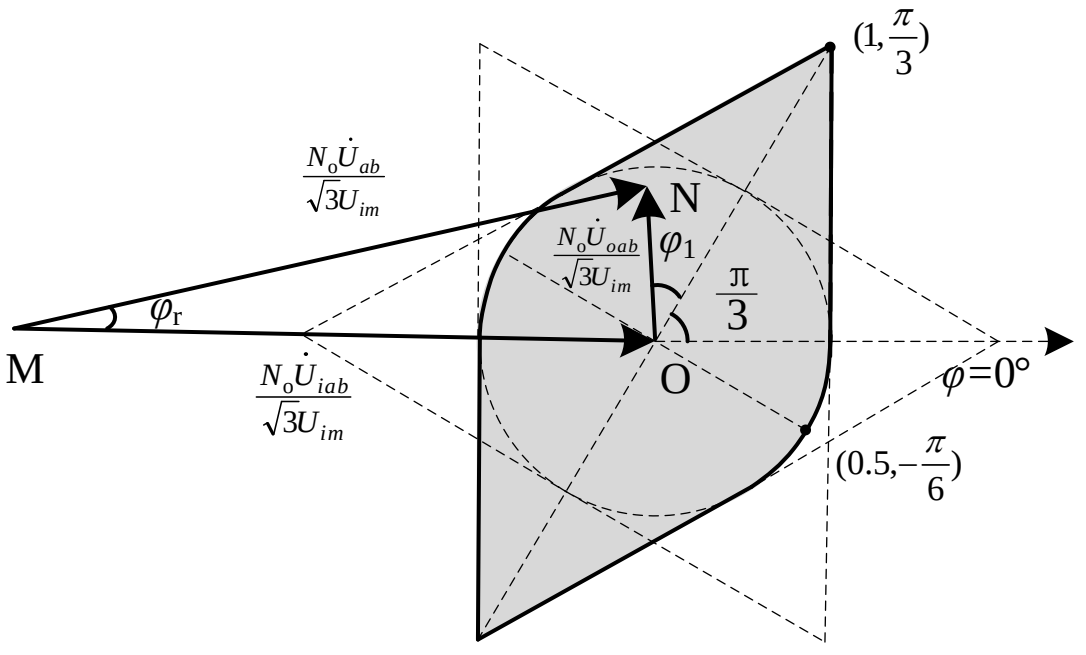

Fig. 9. The total adjustment range of compensation voltage.

*B. Influence of Control Parameters on Adjustment Range*

1)The relationship between adjustable range of $|k_d|$ and $k_0$:

It can be seen from Fig. 8 that the extreme value point of adjustable voltage amplitude is the two intersections of the circle A and the line $\varphi$=0° or 180°, and the adjustment range is the circle based on A as centre and 0.5(1-$k_0$) as the radius. When the adjustable range does not include the origin O, we can get:

$$\begin{cases} |k_d|_{max} = k_0 + \dfrac{k_2}{2} = \dfrac{1+k_0}{2} \\ |k_d|_{min} = k_0 - \dfrac{k_2}{2} = \dfrac{1-k_0}{2} \end{cases} \quad (\frac{1}{3} < k_0 < 1) \tag{18}$$

When the adjustable range includes the origin O, $|OC|_{min}$=0. When -1<$k_0$<0, the situation is similar. Fig. 10 shows the relationship between the regulation range of $|k_d|$ and $k_0$.

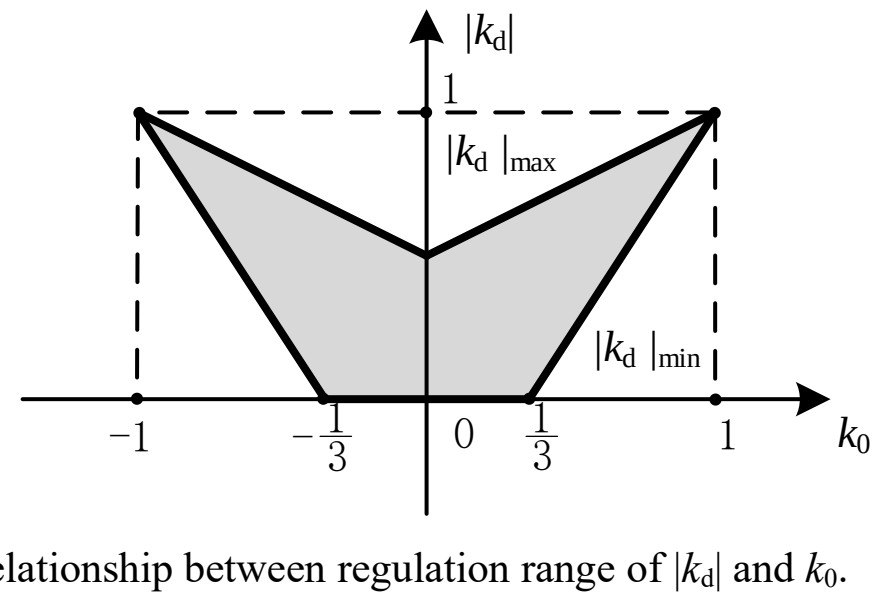

Fig. 10. Relationship between regulation range of $|k_d|$ and $k_0$.

2)The relationship between adjustable range of $\varphi$ and $k_0$:

When the compensation voltage phase takes the extreme value and 0<$k_0$<1, its relationship with $k_0$ is shown in Fig.11. At this time, $k_2$=1-$k_0$, OA=$k_0$, 2AC=1-$k_0$. We can get:

$$\begin{cases} \varphi \in [-\arcsin\dfrac{1-k_0}{2k_0}, \arcsin\dfrac{1-k_0}{2k_0}], \dfrac{1}{3} < k_0 < 1 \\ \varphi \in [0, 2\pi], 0 < k_0 < \dfrac{1}{3} \end{cases} \tag{19}$$

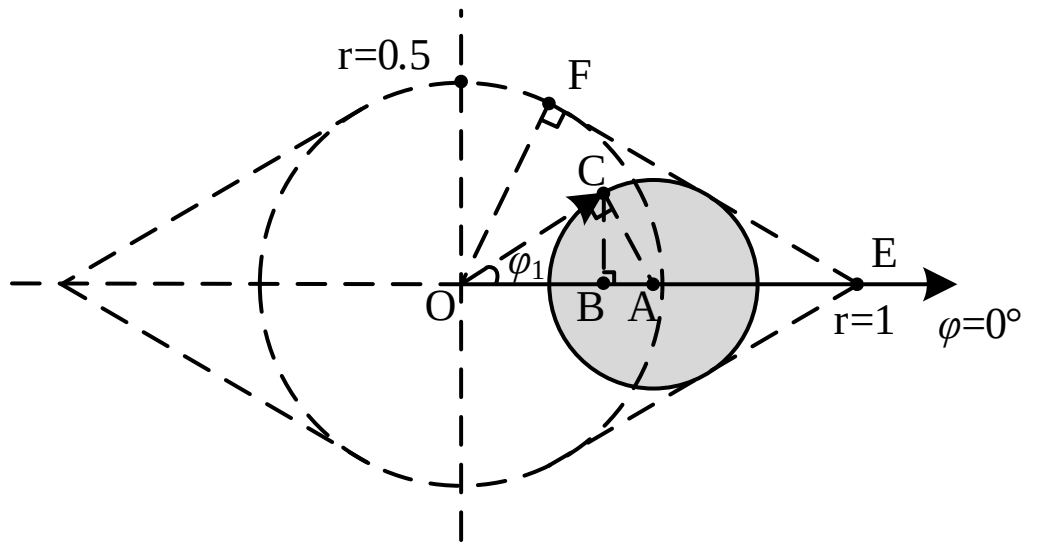

Fig. 11. Compensation voltage regulation range.

Similar when -1<$k_0$<0, and the relationship between the regulation range of $\varphi$ and $k_0$ is shown in Fig. 12.

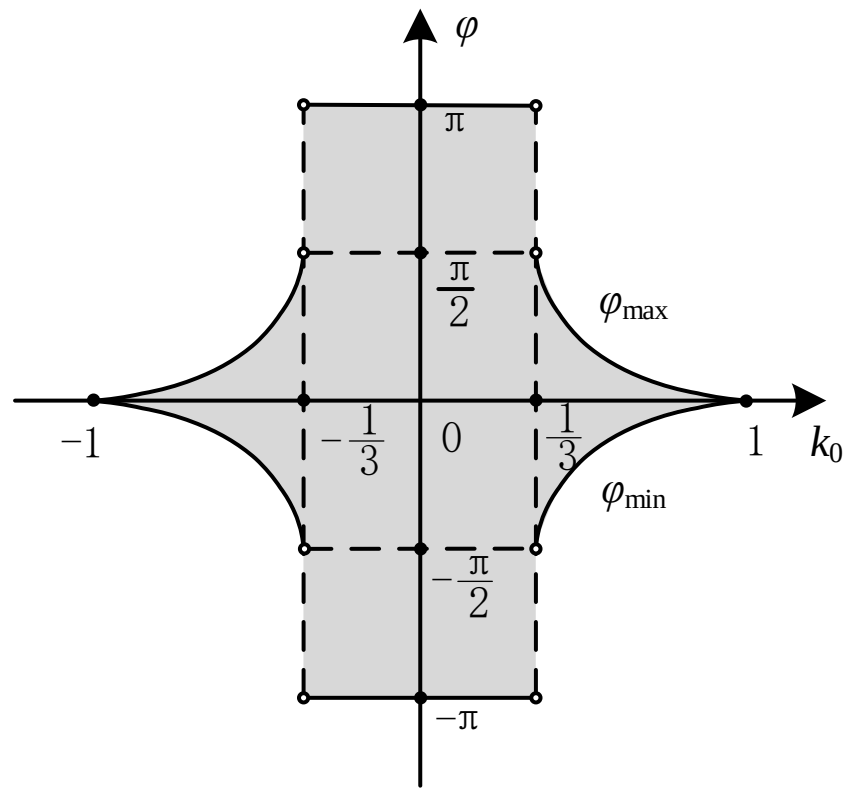

Fig. 12. Relationship between regulation range of $\varphi$ and $k_0$.

*C. Control Strategy*

1) Selection of initial parameters

Known from formula (4), the value of $k_2$ affects the magnitude of the third harmonic voltage in F-DPFC and then it should be taken as small as possible for the stability of grid. A reasonable parameter selection strategy is proposed below and the compensation voltage vector synthesis is shown in the Fig.13.

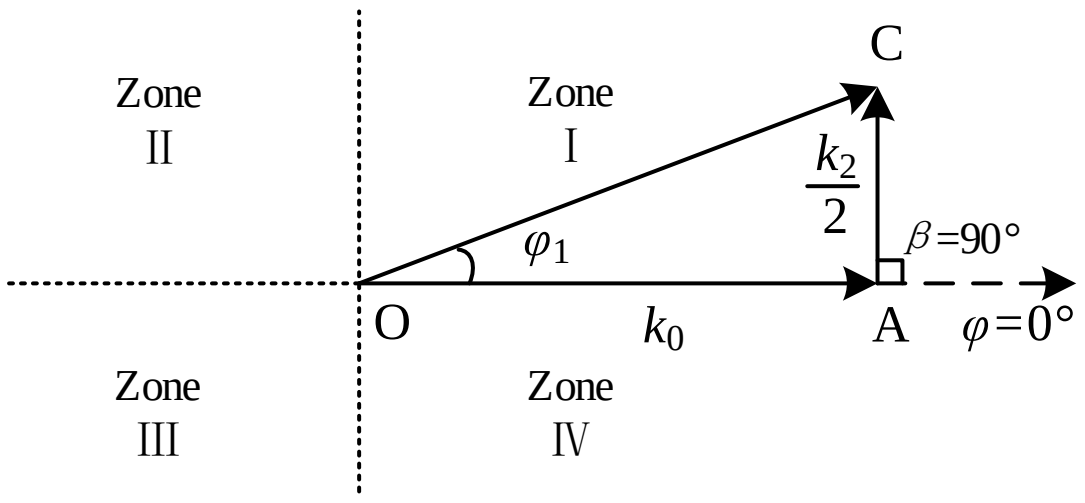

Fig. 13. Compensation voltage vector synthesis under optimal control.

From Fig. 13 and formula (14), in order to minimize the value of $k_2$, the vector $AC$ should be perpendicular to the vector $OA$, that is, $\beta$=90° or -90°. When the output compensation voltage is required to be within the zone I and II, $\beta$=90°; When the output compensation voltage is required to be in the zone III and IV, $\beta$=-90°. Then obtain the value of $k_0$ and $k_2$ based on the amplitude and phase of desired output compensation voltage. The specific process is as follows:

For the convenience of analysis, we assume that $T_o$ are of Δ/Yn11 type connection group and the turn ratio of $T_o$ is 1:1. Taking A-phase as an example, set the required output compensation voltage amplitude and phase as $U_s$ and $\varphi_s$ (where $U_s$ is the amplitude of $u_{oa}$ and $\varphi_s$ is the phase angle of $u_{oa}$ leading $u_{ia1}$). Through the above analysis, one knows that $k_0=U_s\cos(\varphi_s\text{-}30)$ and $k_2=2U_s\sin(\varphi_s\text{-}30)$, when the output compensation voltage is required to be in zone I and II, $\beta$=90°; when the output compensation voltage is required to be in zone III and IV, $\beta$=-90°.

It should be noted that the adjustment range of output compensation voltage synthesized based on the above control strategy is slightly smaller than the total compensation voltage adjustment range analyzed previously, and the range is a rhombus formed by connecting the four intersection point of the edge of total adjustment range and coordinate axis. When the required output adjustment point is outside the range, we can obtain the value range of $k_0$, $k_2$ and $\beta$ based on previous analysis. Due to the small size of this area and there will be a certain margin during operation, we will not discuss it in detail here.

2) Closed-loop Control Strategy

The control object of F-DPFC is the amplitude and phase angle of the output voltage. Firstly, assign the initial value to parameter $k_0$, $k_2$ and $\beta$. During the control process, phase closed-loop is performed first, followed by amplitude closed-loop. $\varphi_{o1}$ which is the phase angle of $u_{oa}$ leading $u_{ia1}$ and $U_{o1}$ which is the amplitude of $u_{oa}$ are obtained by sampling. Then, by comparing them with reference phase angle $\varphi_{ref}$ and voltage amplitude $U_{ref}$, the parameter $k_0$ and $k_2$ are continuous adjusted to generate a new duty cycle signal. If $k_0$ or $k_2$ is 0, the control process is simple because there is only one parameter that needs to be adjusted, so this condition will not be discussed here. One adjustment cycle in detail is as follows:

1. The phase closed-loop: when $\varphi_{o1}>\varphi_{ref}$, if $k_0$>0 and $\beta$=90°, or $k_0$<0 and $\beta$=-90°, $k_2$ decreases, and if $k_0$<0 and $\beta$=90°, or $k_0$>0 and $\beta$=-90°, $k_2$ increases; when $\varphi_{o1}<\varphi_{ref}$, if $k_0$>0 and $\beta$=90°, or $k_0$<0 and $\beta$=-90°, $k_2$ increases, and if $k_0$<0 and $\beta$=90°, or $k_0$>0 and $\beta$=-90°, $k_2$ decreases. When the absolute value of the difference between $\varphi_{o1}$ and $\varphi_{ref}$ is maintained within a small range Δ, the ratio of $k_2$ to $k_0$ is saved at this time and the phase closed-loop is completed.

2. The amplitude closed-loop: when $U_{o1}>U_{ref}$, if $k_0$>0, $k_0$ decreases, and if $k_0$<0, $k_0$ increases, then obtain $k_2$ based on the value of $k_0$ and saved ratio of $k_2$ to $k_0$; when $U_{o1}<U_{ref}$, if $k_0$<0, $k_0$ decreases, and if $k_0$>0, $k_0$ increases, then obtain $k_2$ based on the saved ratio of $k_2$ to $k_0$, until the amplitude closed-loop is finished.

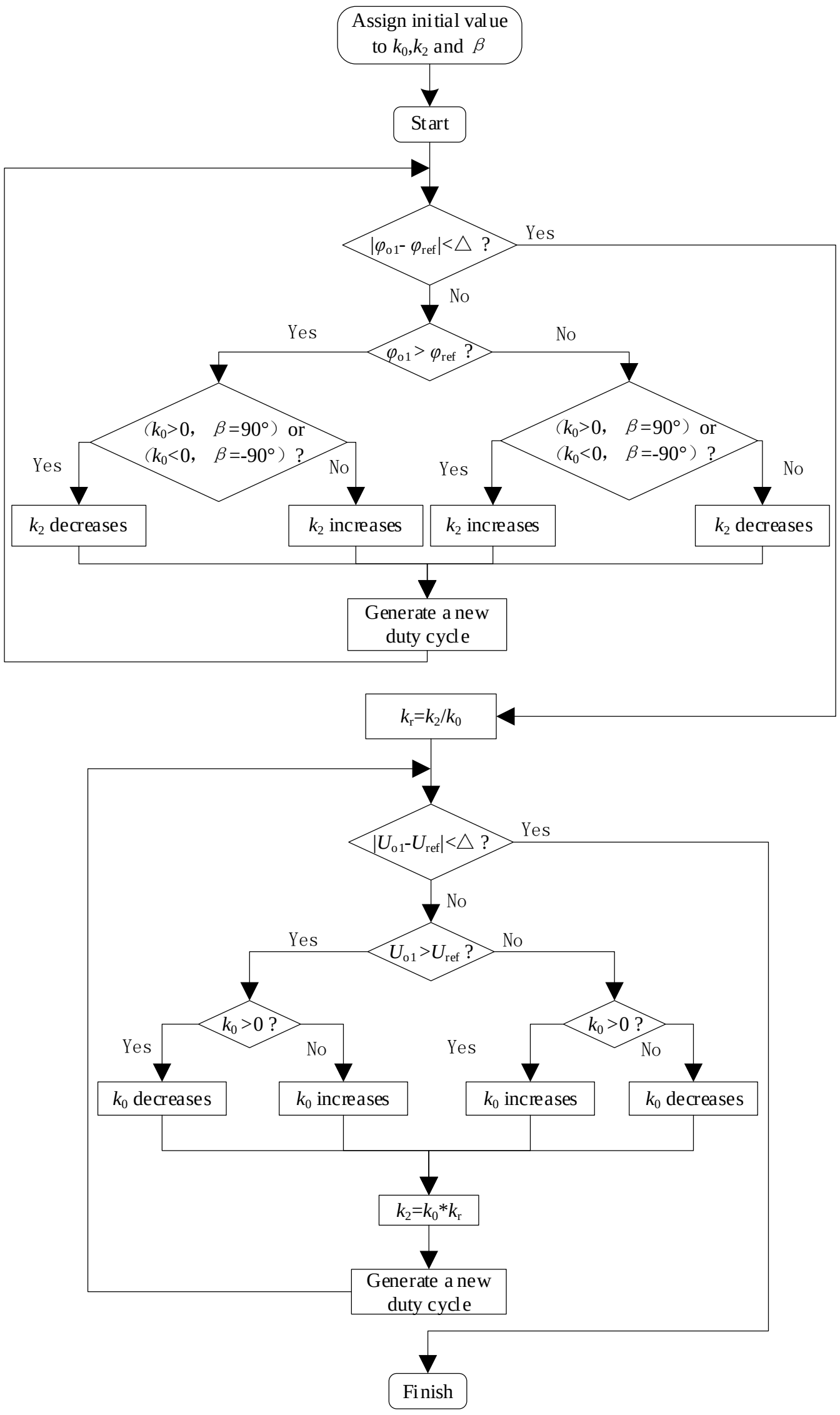

Fig. 14. Closed-loop regulation flowchart.

The flowchart is shown in Fig. 14. It should be noted that start and finish are the beginning and end of a cycle rather than the adjustment process.

## IV. EXPERIMENTAL RESULTS

Through the above theoretical analysis, the prototype as shown in Fig. 15 is established and we conduct experiments on the compensation voltage in different zones. $T_i$ and $T_o$ are of Δ/Yn11 type connection group. To observe the third harmonic voltage, a small capacitor is connected in parallel at the output of each phase transformation unit. Table I lists its specifications.

TABLE I
THE EXPERMENTAL SPECIFICATIONS

| specification | value |
|---|---|
| Amplitude of original grid line voltage: $U_{imL}$ | $200\sqrt{2}$V |
| Frequency of grid voltage: $f$ | 50Hz |
| Switching frequency: $f_s$ | 25kHz |
| Turn ratio of the input transformer $T_i$: $N_i$ | 200/70 |
| Turn ratio of the output transformer $T_o$: $N_o$ | 220/127 |
| Output filter inductance: $L_{fa}=L_{fb}=L_{fc}$ | 0.66mH |
| Output filter Capacitor: $C_{fa}=C_{fb}=C_{fc}$ | 4.4μF |
| Switch device | IRFP460A |
| DSP IC | TMS320F2812 |

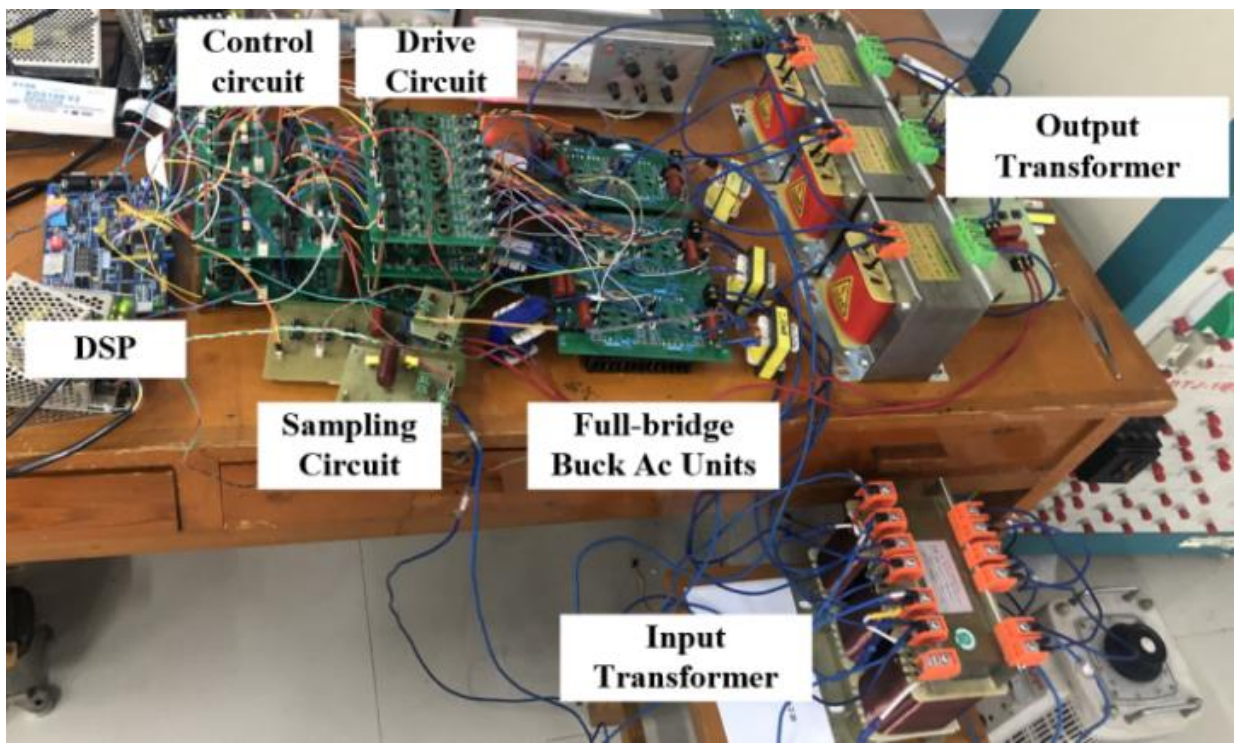

Fig. 15. F-DPFC circuit construction.

In order to minimize the content of the third harmonic in the circuit during operation, the parameter $k_2$ needs to be the minimum value, so the parameter $\beta$ equals to 90° or -90° in this experiment. Since the fluctuation of the power grid is generally within 20%, the input voltage of the full-bridge buck ac unit is set to be about $70\sqrt{2}$V. The parameters of $k_0$ and $k_2$ are continuously adjusted to achieve closed-loop control by using TMS320F2812 chips. The specific experimental results are as follows.

In zone I, when $T_i$ and $T_o$ are of Δ/Yn11 type connection group and the $\varphi_{ref}$ and $U_{ref}$ are respectively 76° and $33\sqrt{2}$V (where $U_{ref}$ is the amplitude of $u_{oa}$ and $\varphi_{ref}$ is the phase angle of $u_{oa}$ leading $u_{ia1}$), Fig. 16 shows the experimental waveforms of F-DPFC.

The input voltage $u_{ia1}$ and outp ut voltage $u_{oa1}$ (between point $E_1$ and point F in Fig. 3) and $u_{oa2}$ (between point $A_3$ and point F in Fig. 3 ) of A-phase buck ac unit in F-DPFC are shown in Fig. 16(a). Among them, $u_{ia1}$ is modulated by the A-phase buck ac unit with a duty cycle of $d_a$, $u_{oa1}$ is a high-frequency pulse sequence, and $u_{ia1}$ is its amplitude envelope curve. Most of the high-frequency components of $u_{oa1}$ are filtered by $L_{fa}$ to obtain $u_{oa2}$. $u_{oa2}$ has not only the fundamental voltage component, but also the third harmonic voltage component and a small number of high-frequency components.

The experimental waveforms of the voltage $u_{oa2}$, $u_{ob2}$ (including the third harmonic component and the fundamental component) and $u_{oa}$ (A-phase output compensation voltage in Fig. 3) are shown in Fig. 16(b). The phase difference between $u_{oa2}$ and $u_{ob2}$ is 120°, of which the third harmonic component is the same and the phase difference of the fundamental component is 120°. Therefore, by offsetting the third harmonic, the output compensation voltage $u_{oa}$ can be obtained, where $u_{oa}=N_o(u_{oa2}-u_{ob2})$.

The experimental waveforms of $u_{iab}$ (the original grid line voltage), $u_{ob}$ (B-phase output compensation voltage in Fig.3), $u_{ab}$ (the regulated grid line voltage) are shown in Fig. 16(c). One knows that $u_{ab}=u_{iab}+u_{oa}-u_{ob}$. With the help of Code Composer Studio Software and DSP simulator, it is easy to observe variables. As can be measured from Fig. 16(c), the phase angle of $u_{ob}$ lags the $u_{iab}$ by 43.5° and we can get the phase angle of $u_{oa}$ leads the $u_{ia1}$ by 76.5° (where $u_{oa}$ leading $u_{ob}$ by 120°, and $u_{ia1}$ and $u_{iab}$ having the same phase), and the amplitudes of $u_{oa}$ is $34\sqrt{2}$V. $u_{ab}$ is leading $u_{iab}$ by 9.6° and has an amplitude of $190\sqrt{2}$V smaller than that of $u_{iab}$ (the measured values are basically the same as the reference values). We can see that under the closed-loop control, the difference between the measured value and the calculated value is very small. The correctness of F-DPFC theoretical analysis is verified.

The experimental waveforms of regulated grid line voltage $u_{ab}$, $u_{bc}$ and $u_{ca}$ (where $u_{ab}=u_{iab}+u_{oa}-u_{ob}$, $u_{bc}=u_{ibc}+u_{ob}-u_{oc}$, $u_{ca}=u_{ica}+u_{oc}-u_{oa}$) are shown in Fig. 16(d), the amplitude of $u_{ab}$, $u_{bc}$ and $u_{ca}$ is $190\sqrt{2}$V. $u_{iab}$ is compensated by $u_{oa}$ and $u_{ob}$ to obtain $u_{ab}$. $u_{ab}$ and $u_{iab}$ have different phases and amplitudes. When the phase difference is 120°, $u_{ab}$, $u_{bc}$ and $u_{ca}$ are positive-sequence symmetrical (usually the original grid line voltage $u_{iab}$, $u_{ibc}$ and $u_{ica}$ are symmetrical, which obviously means compensation phase voltages $u_{oa}$, $u_{ob}$, and $u_{oc}$ must be symmetrical).

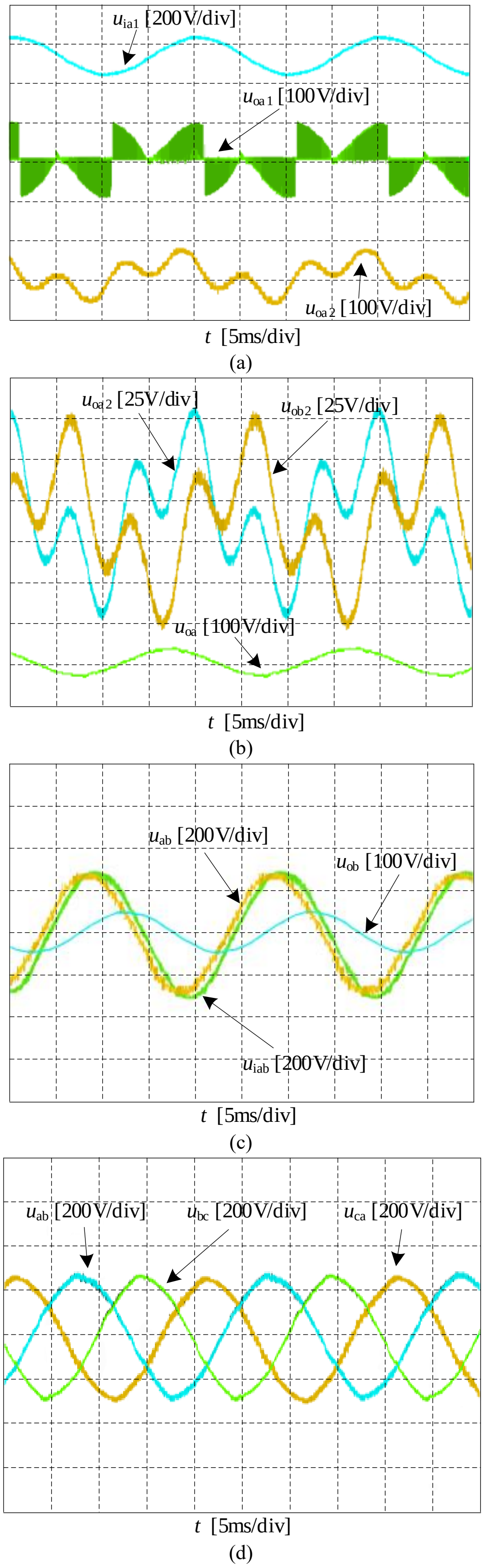

Fig. 16. Experimental waveforms of F-DPFC in zone I when $\varphi_{ref}$=76° and $U_{ref}$=$33\sqrt{2}$V. (a) $u_{ia1}$, $u_{oa1}$, and $u_{oa2}$. (b) $u_{oa2}$, $u_{ob2}$, and $u_{oa}$. (c) $u_{ab}$, $u_{iab}$ and $u_{ob}$. (d) $u_{ab}$, $u_{bc}$, and $u_{ca}$.

In zone II, when $T_i$ and $T_o$ are of Δ/Yn11 type connection group and the $\varphi_{ref}$ and $U_{ref}$ are respectively 170° and $28\sqrt{2}$V. The experimental waveforms of $u_{ia1}$, $u_{oa2}$ and $u_{oa}$ are shown in Fig. 17. The $u_{oa}$ leads the $u_{ia1}$ by 172.5° and the amplitudes of $u_{oa}$ is $28\sqrt{2}$V (the measured values are basically the same as the reference values).

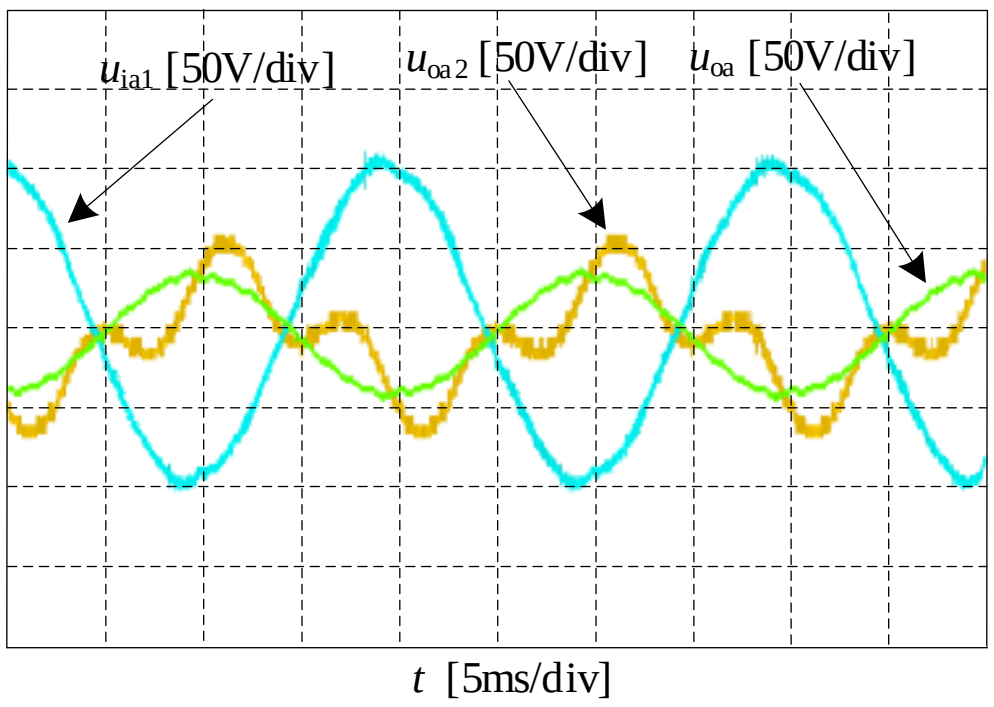

Fig. 17. Experimental waveforms of F-DPFC in zone II when $\varphi_{ref}$=170° and $U_{ref}$=28$\sqrt{2}$V.

In zone III, when $T_i$ and $T_o$ are of Δ/Yn11 type connection group and the $\varphi_{ref}$ and $U_{ref}$ are respectively -120° and 32$\sqrt{2}$V, the experimental waveforms of $u_{ia1}$, $u_{oa2}$ and $u_{oa}$ are shown in Fig. 18. The $u_{oa}$ lags the $u_{ia1}$ by 118.9° and the amplitudes of $u_{oa}$ is 34$\sqrt{2}$V (the measured values are basically the same as the reference values).

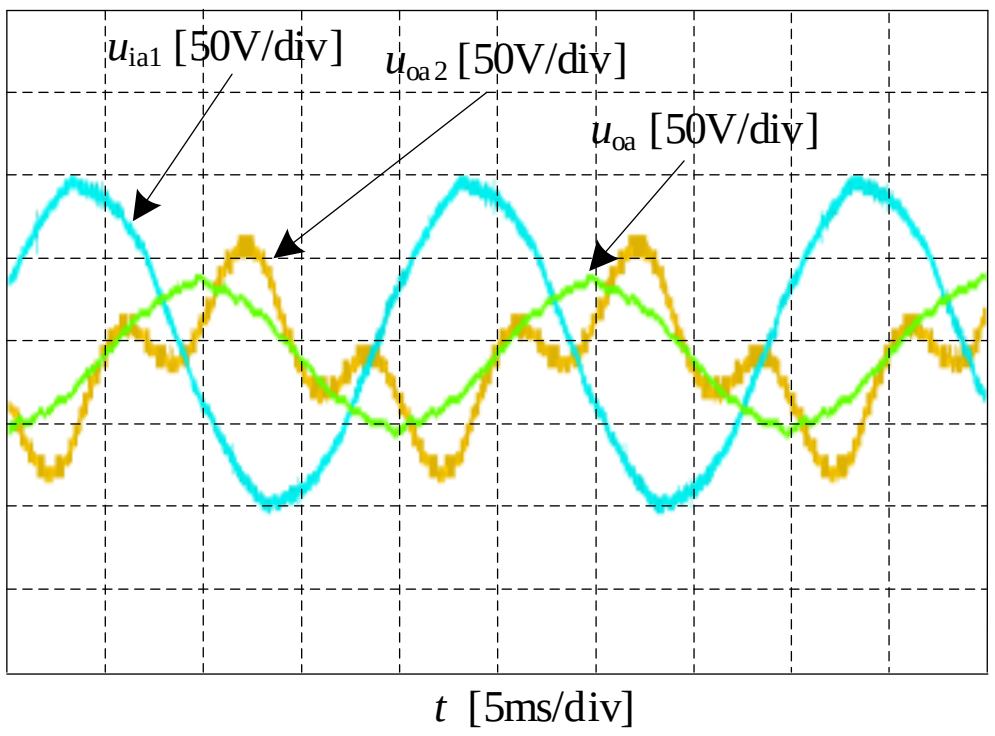

Fig. 18. Experimental waveforms of F-DPFC in zone III when $\varphi_{ref}$=-120° and $U_{ref}$=32$\sqrt{2}$V.

In zone IV, when $T_i$ and $T_o$ are of Δ/Yn11 type connection group and the $\varphi_{ref}$ and $U_{ref}$ are respectively -38° and 26$\sqrt{2}$V, the experimental waveforms of $u_{ia1}$, $u_{oa2}$ and $u_{oa}$ are shown in Fig. 19. The $u_{oa}$ lags the $u_{ia1}$ by 35.3° and the amplitudes of $u_{oa}$ is 25$\sqrt{2}$V (the measured values are basically the same as the reference values).

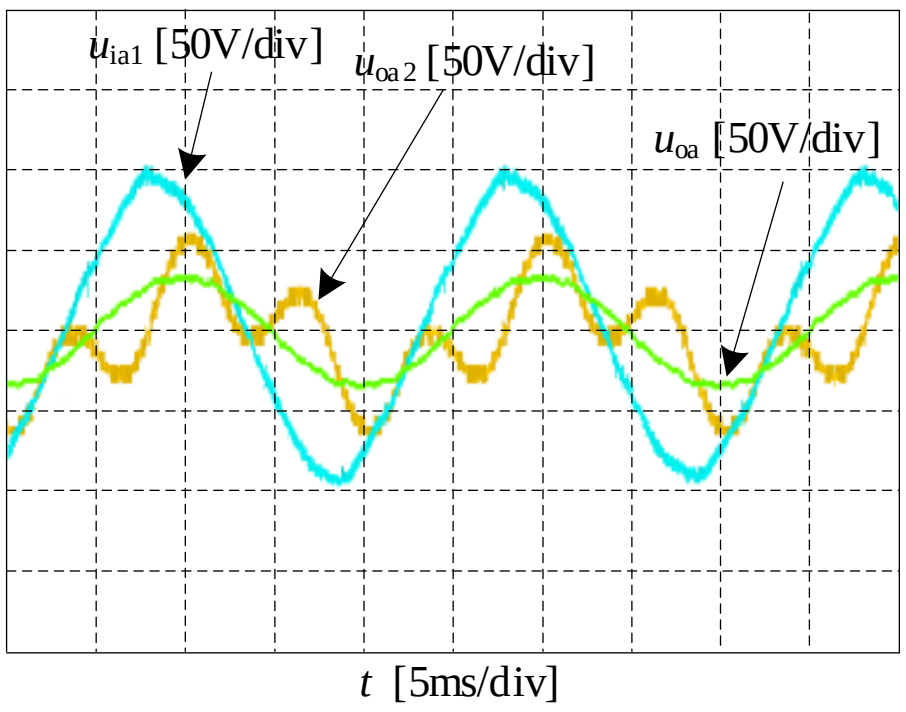

Fig. 19. Experimental waveforms of F-DPFC in zone IV when $\varphi_{ref}$=-38° and $U_{ref}$=26$\sqrt{2}$V.

On the basic of above, eight sets of experiments in four zones are carried out in here, and the experimental results are shown in Table II ($U_{oa}$ and $U_{ia1}$ are the amplitudes of $u_{oa}$ and $u_{ia1}$). $T_i$ and $T_o$ are of Δ/Yn11 type connection group, and we can display all eight groups of data within the compensation voltage regulation range as shown in Fig. 20. It can be seen that the experimental results are almost consistent with the theoretical analysis, which verifies the correctness and feasibility of the theory in this paper.

TABLE II
THE EXPERMENTAL RESULTS

| | Calculated | | Measured(*) | |
|---|---|---|---|---|
| | $\varphi_1$ | $U_{oa}/U_{ia1}$ | $\varphi_{1*}$ | $U_{oa*}/U_{ia1}$ |
| ① | 0° | 0.64 | 0° | 0.63 |
| ② | 52° | 0.49 | 51.1° | 0.47 |
| ③ | 90° | 1 | 90.0° | 0.98 |
| ④ | 143° | 0.41 | 142.3° | 0.41 |
| ⑤ | -174° | 0.63 | -174.9° | 0.64 |
| ⑥ | -141° | 0.43 | -141.8° | 0.44 |
| ⑦ | -80° | 0.43 | -79.7° | 0.43 |
| ⑧ | -47° | 0.34 | -48.1° | 0.33 |

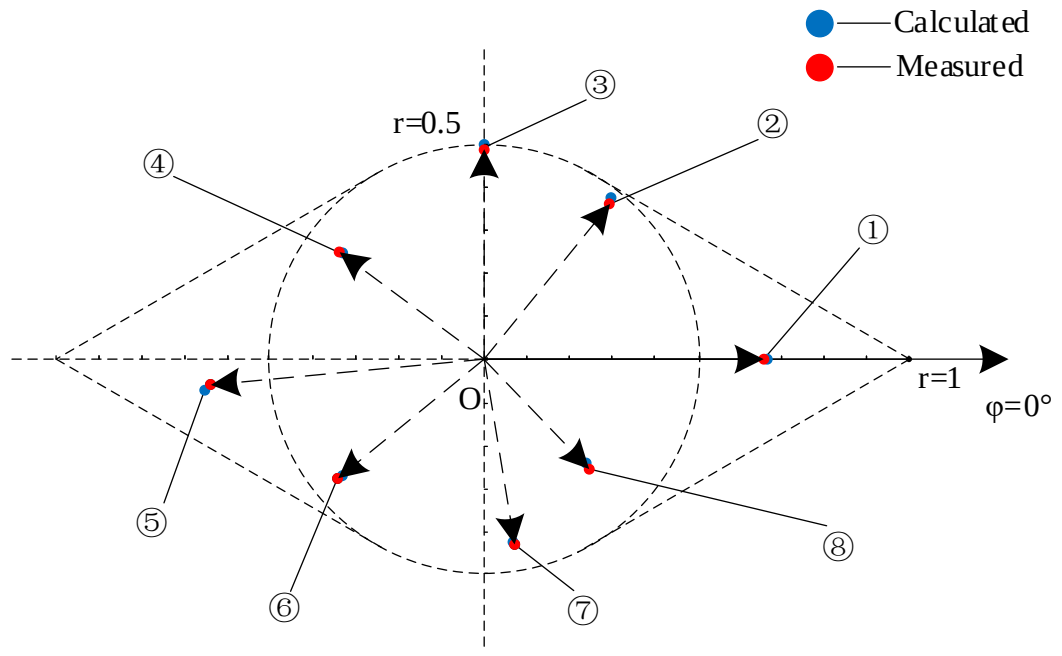

Fig. 20. Experimental results and calculated results.

## V. CONCLUSION

In this paper, a new concept called direct power flow controller with continuous full regulation range (F-DPFC) was proposed. F-DPFC inherits all advantages of DPFC and UPFC and can achieve the regulation of active and reactive power flows in the power grid. F-DPFC simplifies the structure of DPFC and achieves full range continuous regulation. Compared to UPFC, F-DPFC has only one stage conversion circuit and low maintenance costs because dc energy storage elements are not contained. The structure of F-DPFC is simple to achieve modular expansion and enables it to operate under high voltage and power conditions. The operational principle and control strategy of F-DPFC were presented and analyzed in detail, and the principle prototype and experimental results verify the correctness and feasibility of the concept of F-DPFC.

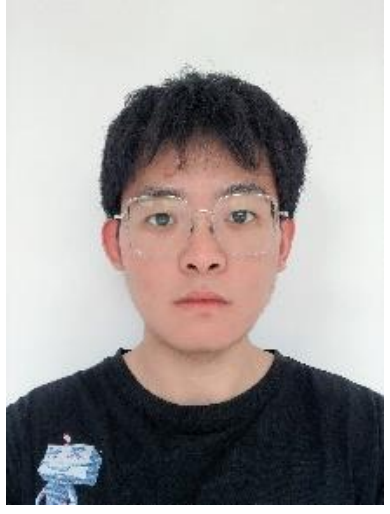

**Yao Chong** was born in Xuzhou, Jiangsu Province, China in 1999. In 2021, he obtained a Bachelor's degree in Automation from Yancheng Institute of Technology in China. He is currently studying for a master's degree in control engineering at Soochow University in Suzhou, China. His main research interests include power electronics conversion technology, especially AC/AC converters for power flow control.

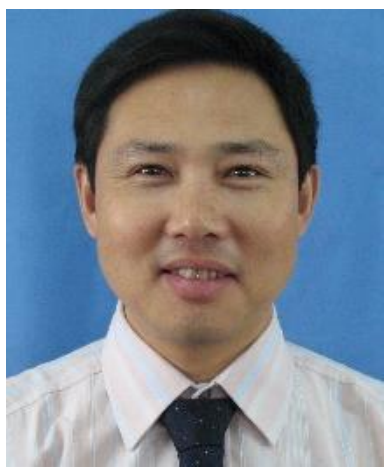

**Youjun Zhang** was born in Huangshan, Anhui Province, China, in 1970. He received the B.S. and M.S. degrees in electrical engineering from Southwest Jiaotong University, Chengdu, China, in 1992, and Nanjing University of Aeronautics and Astronautics, Nanjing, China, in 2002, respectively, where he received the Ph.D. degree in electrical engineering in 2014.

In 2004, he joined School of Mechanical & Electrical Engineering, Soochow University, where he became a professor in 2018. He has been a co-researcher with Jiangsu Key Laboratory of Spectral Imaging & Intelligent Sense and Ministerial Key Laboratory of JGMT, Nanjing University of Science and Technology, Nanjing, China, since 2014 and 2016, respectively. His main research interests include ac/ac converters for power flow control, multilevel converters, and dc/ac converters.